\documentclass[aps,pra,twocolumn]{revtex4-2}
\usepackage{amsmath,epsfig,mathrsfs,graphicx,amssymb,color}
\usepackage[T1]{fontenc}
\usepackage{t1enc}
\usepackage{hyperref}
\usepackage{array}
\usepackage{booktabs}

\def\be#1\ee{\begin{equation}#1\end{equation}}
\newcommand{\ba}{\begin{eqnarray} }
\newcommand{\ea}{\end{eqnarray} }

\usepackage[cp1250]{inputenc}   % pozwala wprowadzaÄ‡ polskie znaki bezpoĹ›rednio z klawiatury
\usepackage{amsmath}            % AMS-LaTeX (AMS-Math)
\usepackage{amsfonts}           % ekstra fonty matematyczne
\usepackage{amssymb}            % ekstra symbole matematyczne
\usepackage{enumerate,graphicx}
\usepackage{array}
\usepackage{booktabs}

\usepackage{physics}

\def\mb{\begin{pmatrix}}
\def\me{\end{pmatrix}}
\def\be#1\ee{\begin{equation}#1\end{equation}}

\begin{document}

\title{Device-independent dimension leakage null test on qubits at low operational cost}

\author{Tomasz Rybotycki$^{1,2,3}$}
\author{Tomasz Bia{\l}ecki$^{4}$}
\author{Josep Batle$^{5,6}$}
\email{jbv276@uib.es, batlequantum@gmail.com}
\author{Adam Bednorz$^{4}$}

\email{Adam.Bednorz@fuw.edu.pl}

\affiliation{$^1$Systems Research Institute, Polish Academy of Sciences, 6 Newelska Street, PL01-447 Warsaw, Poland}
\affiliation{$^2$Nicolaus Copernicus Astronomical Center, Polish Academy of Sciences, 18 Bartycka
Street, PL00-716 Warsaw, Poland
}
\affiliation{$^3$Center of Excellence in Artificial Intelligence, AGH University,
30 Mickiewicza Lane, PL30-059 Cracow, Poland
}
\affiliation{$^4$Faculty of Physics, University of Warsaw, ul. Pasteura 5, PL02-093 Warsaw, Poland}
\affiliation{$^5$Departament de F\'isica and Institut d'Aplicacions Computacionals de Codi Comunitari (IAC3), Campus UIB, E-07122 Palma de Mallorca, Balearic Islands, Spain}
\affiliation{$^6$CRISP -- Centre de Recerca Independent de sa Pobla, 07420 sa Pobla, Balearic Islands, Spain}

\begin{abstract}
We construct a null test of the two-level space of a qubit, which is both device independent and needs a small number of different experiments.
We demonstrate its feasibility on IBM Quantum, with most qubits failing the test by more than 10 standard deviations. The robustness of the test
against common technical imperfections, like decoherence and phase shifts, and supposedly negligible leakage, indicates that the origin of deviations is beyond known effects.

\end{abstract}

\maketitle

\section{Introduction}

Efficient error correction in quantum computations relies on the assumption that the qubit is a qubit, i.e. a two-level space.
Otherwise, unidentified leakage to extra space counts as an irrecoverable error, which may accumulate in long operational sequences.

To verify the qubit's space one can use a dimension witness. The witness
 is a function of several quantities (probabilities) measured in a certain protocol. Device independence allows to involve untrusted operations,
 while the witness checks if they do not take the system to extra states.

The  typical protocol is a prepare-and-measure scenario \cite{gallego}, consisting of a sequence of two operations
which are formally called preparation and measurement. Each one can be independently chosen from a restricted set. Such linear witnesses were based
on inequalities, and have been tested experimentally \cite{hendr,ahr,ahr2,dim1} but they are useless if the contribution from extra states is small. In the latter case, a better choice is a nonlinear witness equal to a determinant of probability matrix, equal to zero, up to statistical error, when the expected dimension is matched \cite{dim, chen,bb22, epj}. For a qubit it requires minimally $20$ combinations to be measured. One can reduce it to 8 combinations \cite{pra24}, involving a repeated operation. However, the usage of a unitary operation gives the degenerate adjugate probability matrix, which complicates the search for potential deviations as they become very small. 

Here, we propose a modified witness using also a single repeated operation but we allow two independent initial states.
The witness remains a determinant but its bounds in general dimension require partially numerical derivation.
Instead of a simple Toeplitz matrix with elements depending on the multiplicity of the repeated operation (also known as the method of delays) \cite{wgp,leak}, it consists of two blocks for two initial states.
We show that the minimal number of independent experiments (operational cost) is 11 in the above construction, and the test can be easily implemented in IBM Quantum devices, \emph{ibm\_brisbane}, and \emph{ibm\_sherbrooke},
in our case. Device independence means here that the witness is zero for an arbitrary state preparation, the applied operation and 
the measurement process, which can be imperfect or unknown (e.g. due to noise or systematic errors), on condition that they are restricted to the two-level space the sequence and the identity of the repeated operation is preserved.
The failure of many tested qubits by 10 or more standard deviations shows that there is an unspecified leakage to extra space, even if the operations are not perfect.

\begin{figure*}
	\includegraphics[scale=.7]{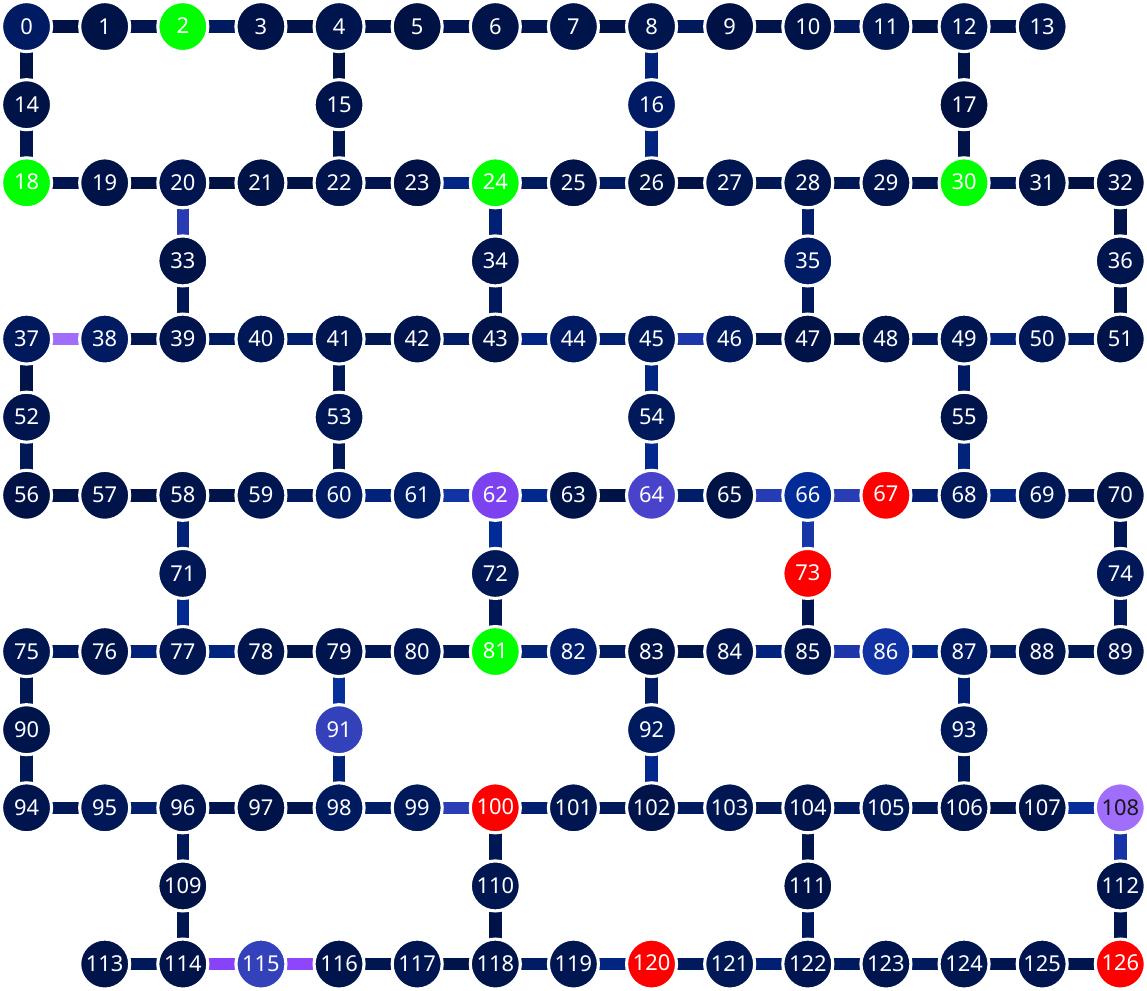}
	\caption{The topology of IBM Quantum \emph{ibm\_brisbane}. We have highlighted the most
		faulty qubits tested in  April 2024 -- red and
		green -- the rest of the qubits that have passed the test.
		Two-qubit echoed cross-resonance gates, unused in the test, connect the
		qubits.}
	\label{brisq}
\end{figure*}

\begin{figure*}
	\includegraphics[scale=.7]{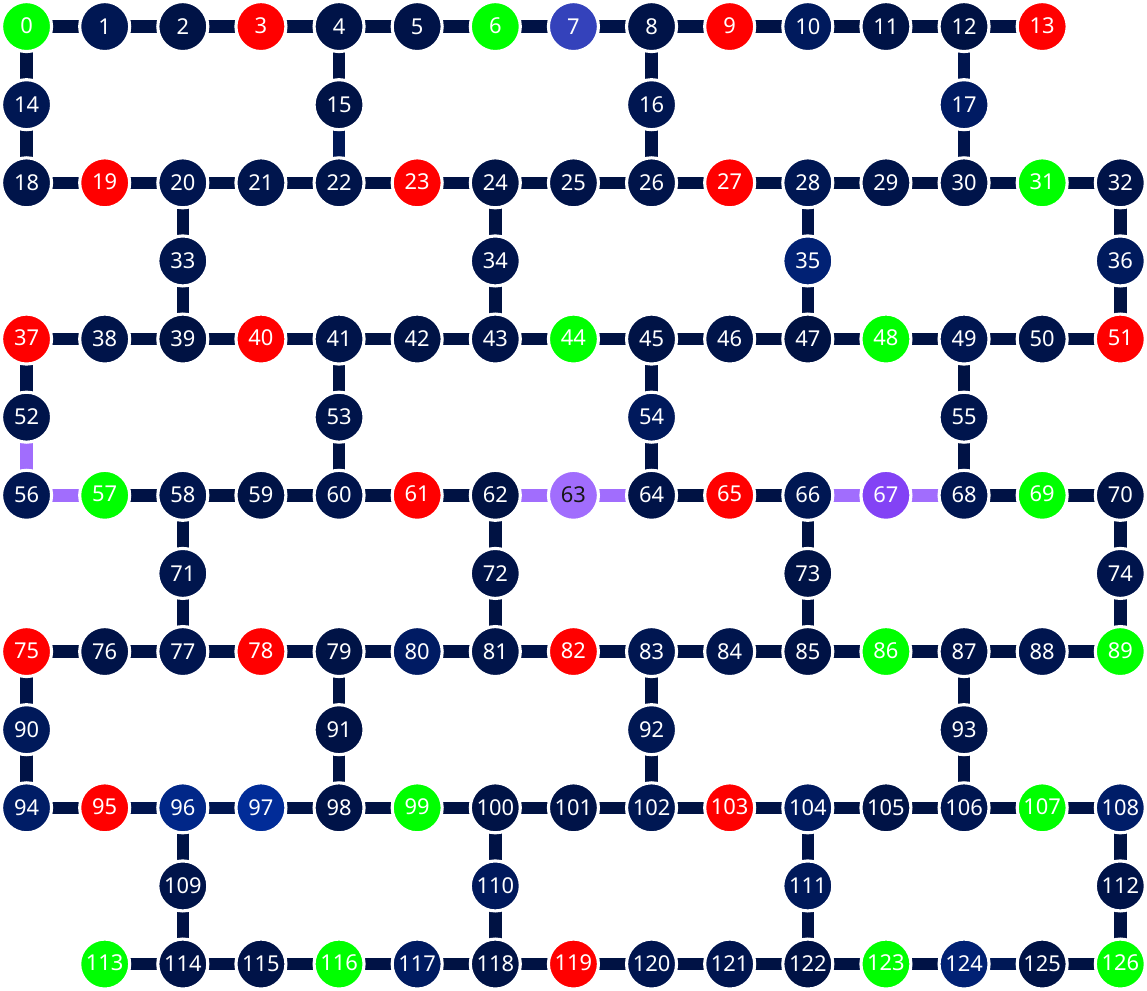}
	\caption{The topology of IBM Quantum \emph{ibm\_sherbrooke}. We have highlighted the most
		faulty qubits tested in May 2024 -- red and
		green -- the rest of the qubits that have passed the test.
		Two-qubit echoed cross-resonance gates, unused in the test, connect the
		qubits.}
	\label{sherq}
\end{figure*}

\section{Construction of the witness}

Analogously as in previous construction \cite{wgp,leak,pra24}, the probability of the outcome determined by the measurement operator $\hat{1}\geq \hat{M}\geq 0$ after $n$ subsequent quantum operations $\mathcal E$ 
(superoperator, a linear map on an operator)
on the initial state $P_i$, $i=1,2$ (in our notation on the left, meaning the time order from the left to the right) is
\be
p_{in}=\mathrm{Tr}\hat{P}_i\mathcal E^{n}\hat{M}.\label{mrp}
\ee
 The operation
must preserve normalization, i.e.  $\mathcal E \hat{1}=\hat{1}$ for identity $\hat{1}$, and complete positivity, i.e. decomposable $\mathcal E\hat{M}=\sum_j\hat{K}^\dag_j\hat{M}\hat{K}_j$.
Suppose the linear space of possible measurements is $\leq D$, including the identity.
From the Cayley-Hamilton theorem, the characteristic polynomial $w(\mathcal E)$ is of degree $\leq D$, divisible by $\mathcal{E}-1$ since one of the eigenvalues is $1$.
We construct the witness $W=\det\;p$ for the matrix
\be
p=\begin{pmatrix}
p_{10}&p_{11}&p_{12}&p_{20}&p_{21}\\
p_{11}&p_{12}&p_{13}&p_{21}&p_{22}\\
p_{12}&p_{13}&p_{14}&p_{22}&p_{23}\\
p_{13}&p_{14}&p_{15}&p_{23}&p_{24}\\
1&1&1&1&1
\end{pmatrix}.
\label{wit1}
\ee
which involes 11 independntly measured probabilities.
If the operations, and their concatenations, remain in the two level space, $d=2$, $D=4$ (matrices $d\times d$), the $5\times 5$ matrix of  preparations $\times $ measurements
has the rank 4 so the determinant must vanish, see also \cite{bb22,epj,pra24}. Note that reduction to a single preparation $P_1=P_2$ at $d=2$
will also make the determinant vanish but the adjugate matrix vanishes, too, which makes more complicated error analysis \cite{pra24}.
The absolute maximum is 3 (equal the maximal determinant of a $0-1$ matrix) for $p_{10}$,$p_{12}$,$p_{13}$,$p_{15}$,$p_{20}$,$p_{24}$ equal to $1$ and $p_{11}$,$p_{14}$,$p_{21}$,$p_{22}$,$p_{23}$ equal to 0,
which can be achieved in classical systems of dimension 9 (with the operation of cyclic shift by 1, initial states at $5$ and $1$, respectively,
 and the measurement at $1,5,7,8$). To find the maximum value at quantum dimension 3 one has to maximize over all possible operations $\mathcal E$, initial states, and the measurement. It turns out that the optimal set (the same in the real and complex space) is defined by the unitary $\mathcal E\hat{M}=\hat{U}^2\hat{M}\hat{U}^{2\dag}$, with
\be
\hat{U}=\begin{pmatrix}
\cos\phi&-\sin\phi&0\\
\sin\phi&\cos\phi&0\\
0&0&1\end{pmatrix},
\ee
in the basis $|1\rangle$, $|2\rangle$, $|3\rangle$,
and the initial states $\hat{P}_1=\hat{U}^{-5}|\psi_1\rangle\langle\psi_1|\hat{U}^{5}$ and $\hat{P}_2=\hat{U}^{-4}|\psi_2\rangle\langle \psi_2|\hat{U}^4$, and $\hat{M}=|\psi_3\rangle\langle\psi_3|$
with normalized $|\psi_i\rangle=a_i|1\rangle+b_i|3\rangle$ for real $a_i,b_i$. The numerical maximum of $W$ over a 4-dimensional space of parameters is $\simeq 0.5259128034146499$.

\section{Error analysis}

Analogously as in the previous works \cite{pra24,epj}, we assume independence of experiments. The nonlinearity of the witness, and the fact that the same probabilities
appear many times in the matrix, leads to a cumbersome expression for the error for $N$ trials.
The dominant source of errors is finite statistics. For a witness, which is a function of binary probabilities
in independent experiments, $W(\{p\})$, we obtain experimentally $\tilde{p}=n/N$, the actual frequency of $n$ results $1$ out of $0/1$
for $N$ repetitions. Denoting the theoretical $p$ is the limit at $N\to \infty$ and $\delta p=\tilde p-\delta p$, we have
\begin{equation}
\langle \delta p_a\rangle=0,\;\langle\delta p_a\delta p_b\rangle=\frac{p_a(1-p_a)}{N}\delta_{ab},
\end{equation}
for experiments $a$ and $b$.
At large $N$ we can expand
\begin{equation}
\delta W\simeq\sum_a\frac{\partial W}{\partial p_a}\delta p_a,
\end{equation}
for $a=10,11,12,13,14,15,20,21,22,23,24$.
Combining the above equations we can express the error
\begin{align}
&N\langle(\Delta W)^2\rangle=N\sigma^2\simeq b_{10} \mathrm{A}_{11}^2+b_{11}(\mathrm{A}_{12}+\mathrm{A}_{21})^2\nonumber\\
&
+b_{12}(\mathrm{A}_{13}+\mathrm{A}_{22}+\mathrm{A}_{31})^2+b_{13}(\mathrm{A}_{14}+\mathrm{A}_{23}+\mathrm{A}_{32})^2\nonumber\\
&
+b_{14}(\mathrm{A}_{24}+\mathrm{A}_{32})^2+b_{15} \mathrm{A}_{33}^2+b_{20}\mathrm{A}_{14}^2+b_{21}(\mathrm{A}_{24}+\mathrm{A}_{15})^2\nonumber\\
&
+b_{22}(\mathrm{A}_{34}+\mathrm{A}_{25})^2+b_{23}(\mathrm{A}_{44}+\mathrm{A}_{35})^2+b_{24}\mathrm{A}_{45}^2,
\end{align}
with  $b_a=p_a(1-p_a)$ and the adjugate matrix $\mathrm{A}=\mathrm{Adj}\;p$ (matrix of minors of $p$, with a given row and column crossed out, and then transposed, with the minus sign for a different parity of indices of the column and row). Note that the identity 
$p^{-1}\det p=\mathrm{Adj}\;p$ makes no sense here as $\det p=0$ in the limit $N\to\infty$.

\begin{figure}[t!]
	\includegraphics[scale=.8]{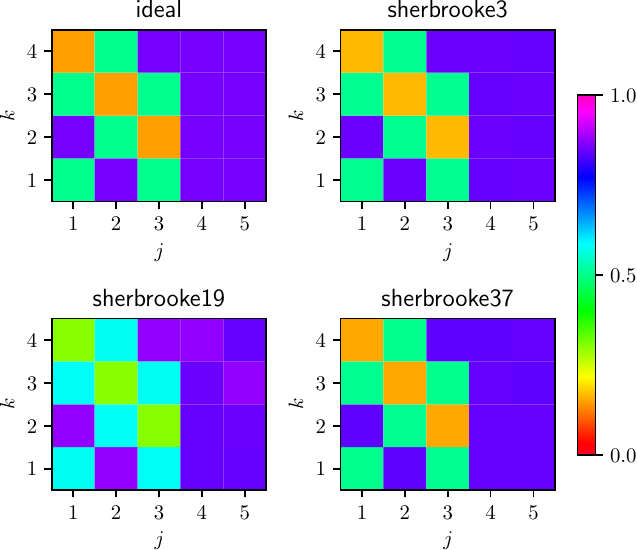}
	\caption{The probabilities $p_{kj}$ for the test,
		ideal,  and the most faulty qubits,  \emph{ibm\_sherbrooke} qubit 3, 19, 37 (May 2024). }
	\label{dmesh}
\end{figure}

\begin{figure}[t!]
	\includegraphics[scale=1]{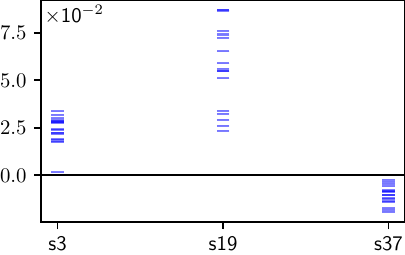}
	\caption{The results of the witness for individual jobs for the most faulty qubits,  \emph{ibm\_sherbrooke} qubit 3, 19, 37 (May 2024). Each blue line represents the value of $W$ calculated for an individual job.}
	\label{dscat}
\end{figure}

\section{Test on IBM Quantum}

The native gate realized by IBM Quantum is the  $\pi/2$ rotation on the Bloch sphere, in $\ket 0$, $\ket 1$ space,
\begin{equation}
	\hat{S}=\frac{1}{\sqrt{2}}\begin{pmatrix}
		1&-i\\
		-i&1\end{pmatrix}\label{smat}.
\end{equation}
The rotation axis can be itself rotated for a given angle $\theta$, which is realized with the above gate $S$ and two auxiliary gates $Z(\theta)$ (phase shifts), 
\begin{equation}
	\hat{S}_\theta=\hat{Z}^\dag_\theta \hat{S}\hat{Z}_\theta ,\:
	\hat{Z}_\theta=
	\begin{pmatrix}
		e^{-i\theta/2}&0\\
		0&e^{i\theta/2}\end{pmatrix}.
\end{equation}
For the test on IBM Quantum we have chosen operation $\mathcal E\hat{M}=\hat{S}^\dag\hat{M}\hat{S}$,the initial state $\hat{P}_1=|1\rangle\langle 1|$
and $\hat{P}_2=\hat{S}_{\pi/2}|1\rangle\langle 1|\hat{S}^\dag_{\pi/2}$ with the measurement $\hat{M}=\hat{S}_{-\pi/4} |0\rangle\langle 0|\hat{S}_{\pi/4}$.
It is important that our choice tests the full complex space of a qubit. Note that reducing to preparations and measurements on a single Bloch circle makes the adjugate matrix $A$ equal to $0$. Thus, it would complicate our error analysis, with additional second-order minors involved, see \cite{pra24}.
Taking preparations and measurements out of a single Bloch circle, we test the full 4-dimensional qubits space, with the non-zero adjugate matrix
\begin{equation}
A=\frac{\sqrt{2}}{8}\begin{pmatrix}
0&0&0&0&0\\
0&0&0&0&0\\
0&0&0&0&0\\
-1&1&-1&1&0\\
1&-1&1&-1&0
\end{pmatrix}.
\end{equation}

We have run the test on two 127-qubit devices, \emph{ibm\_brisbane} and \emph{ibm\_sherbrooke}, in April and May 2024, respectively.
In the first case, we have randomly chosen 10 qubits, and have run 32 jobs  at 10000 shots, with 20 repetitions in each job of each of 11 tests. It gives the total number of repetitions $N=32\cdot 10000\cdot 20=64\cdot 10^5$. 
In the second case we have chosen 32 qubits, and have run 16 jobs at 8000 shots, with 20 repetitions, giving $N=256\cdot 10^4$. 
The random shuffling of the repeated test within each job prevents possible residual memory.
By imposing the absence of software optimization of the gate sequence, the actual gate pulses are performed on the indicated qubits.
The operations on different qubits can be applied independently. To avoid crosstalks, the chosen qubits are at least 3 qubits away from the other tested ones.
The chosen qubits are depicted in 
Figs. \ref{brisq} amd \ref{sherq}. We have calculated the witness in two ways: (i) averaging all probabilities over the jobs and calculating the witness for the averaged 
probability, (ii) calculating the witness for probabilities for individual jobs, and then averaging the obtained witnesses. These two ways of calculations are to 
avoiding the effect of suddenly changed calibration and strong sensitivity of the gates to variable external conditions.
Both tests have been run in single days which excludes large daily calibration changes.
We summarize the results of the tests in Table \ref{data}. It turns out that about half of
qubits fail the test. Especially the second test revealed qubits failing by more than 60 standard deviations. We depicted the actual matrix $p$ and results from individual jobs for the most
failing qubits, see. Fig. \ref{dmesh} and \ref{dscat}. One can find also the $p-$value, i.e. the probability of not rejecting the null hypothesis of the two-level space and the remaining assumptions of the test, taking the doubled tail (events $>|z|$) of the centered normal distribution with standard deviation $\sigma$,
\begin{equation}
p(z)=2\int_z^\infty e^{-z^2/2\sigma^2}/\sqrt{2\pi\sigma^2}=\mathrm{erfc}\;(z/\sqrt{2}\sigma)
\end{equation}
applied to $z=W$, as the distribution is close to Gaussian at large number of trials.
The extremal cases are  \emph{ibm\_brisbane} qubit 126 and \emph{ibm\_sherbrooke} qubit 19, with $W$ at $7.9$ and $71$ standard deviations, and $p$-values $4\cdot 10^{-15}$ and $< 10^{-1000}$, respectively. 
To explain the nonzero value of $W$ by a leakage leaing to small corrections to probabilities i.e. $p\to p+\delta p$, one can estimate
$
\delta W=\mathrm{Tr} A\delta p$ for the adjugate matrix $A$.
 Note that the estimated leakage out of computational space is below $10^{-6}$, see \cite{leak2} and Appendix, while the adjugate matrix elements remain $<1$,
which cannot explain the deviation $>10^{-3}$. For the ideal $S$ gates, corresponding to $\pi/2$ rotation on the Bloch sphere, $p_{10}=p_{14}$,
$p_{11}=p_{15}$, and $p_{20}=p_{24}$. As a sanity check, we extracted the differences from the collected data, Table \ref{data}, with most of them corresponding
to the error $\lesssim 10^{-3}$ per $S$ gate. 
We stress that the test is robust against common sources of error such as decoherence and relaxation within the computational space.
The data and scripts are available in a public repository \cite{zen}.

\begin{table*}

	\begin{tabular}{c|*{17}{c}}
		\toprule
		qubit (brisbane) & 2&18&24&30&{\bf 67}&81&{\bf 73}&{\bf 100}&{\bf 120}&{\bf 126}\\
		drive freq.
		 [GHz]&4.61&4.788&5.101&4.733&5.113&4.93&4.98&4.905&4.837&4.908\\
		gate error $[10^{-4}]$&2.8&1.9&1.4&2.4&4.2&1.9&1.7&1.7&3.1&2\\
		$W^{i}[10^{-5}]$& -27&-0.9&-7.4&-5&{\bf -46}&-0.1&{\bf -60}&{\bf -47}&{\bf -44}&{\bf -69}\\
		$\sigma^i[10^{-5}]$&8.8&8.3&8.1&8.8&{\bf 5.9}&3&{\bf 8.7}&{\bf 8.9}&{\bf 7.1}&{\bf 8.9}
		 \\
		$W^{ii}[10^{-5}]$&-27&-1.4&-7.5&-4.5&{\bf -46}&0.2&{\bf -60}&{\bf -47}&{\bf -44}&{\bf -69}\\
		$\sigma^{ii}[10^{-5}]$&8.8&8.3&8.1&8.8&{\bf 5.9}&3&{\bf 8.7}&{\bf 8.9}&{\bf 7.1}&{\bf 8.9} \\
		$p_{10}-p_{14}[10^{-4}]$&-9.8&32&6.4&-19&32&3.9&58&44&-1.4&15\\
		$p_{11}-p_{15}[10^{-4}]$&3.3&-30&-16&-31&21&-29&1.6&-26&3.4&-8.8\\
		$p_{20}-p_{24}[10^{-4}]$&-8.1&14&-11&30&-0.6&-5.7&-1.4&-3.7&-47&-59\\
		\midrule
		qubit (sherbrooke) & 0&{\bf 3}&6&{\bf 9}&{\bf 13}&{\bf 19}&{\bf 23}&{\bf 27}&31&{\bf 37}&{\bf 40}&44&48&{\bf 51}&57&{\bf 61}\\
		drive freq. [GHz]&4.636&4.747&4.9&4.638&4.557&4.821&4.758&4.68&5.058&4.564&4.706&4.868&4.707&4.767&4.835&4.902 \\
		gate error $[10^{-4}]$&2.6&2.7&1.7&6.3&1.4&3.2&1.5&3.2&4.5&4.2&2.6&1.6&1.6&2.5&2.9&2.7 \\
		$W^{i}[10^{-4}]$&3.1&{\bf 23}&0.0&{\bf 19}&{\bf 22}&{\bf 52}&{\bf 11}&{\bf 11}&5.7&{\bf -10}&{\bf 13}&-1.8&3.8&{\bf 15}&7&{\bf 12} \\
		$\sigma^i[10^{-4}]$&1.3&{\bf 1.3}&1&{\bf 1}&{\bf 1.1}&{\bf 0.7}&{\bf 1.3}&{\bf 1.4}&1.4&{\bf 1.4}&{\bf 1.4}&1.4&1.3&{\bf 1.4}&1.3&{\bf 1.3}
		 \\
		$W^{ii}[10^{-4}]$&3.1&{\bf 24}&0.0&{\bf 22}&{\bf 22}&{\bf 55}&{\bf 11}&{\bf 11}&5.7&{\bf -10}&{\bf 13}&-1.7&3.8&{\bf 15}&7&{\bf 12}\\
		$\sigma^{ii}[10^{-4}]$&1.3&{\bf 1.3}&1&{\bf 1.1}&{\bf 1.1}&{\bf 0.8}&{\bf 1.3}&{\bf 1.4}&1.4&{\bf 1.4}&{\bf 1.4}&1.4&{\bf 1.3}&1.4&1.3&{\bf 1.3}\\
		$p_{10}-p_{14}[10^{-4}]$&1.6&-3.9&-4.9&5&-3.2&-3.3&2.5&-5.2&-8.4&3.8&0.5&-0.9&-0.4&4.3&3.9&6.4\\
		$p_{11}-p_{15}[10^{-4}]$&-0.3&-5.6&3.9&11&-0.7&1.5&-1.1&2.6&1.1&-2.3&0.9&-1.4&-4.6&-5.1&-1.3&-1\\
		$p_{20}-p_{24}[10^{-4}]$&-3.1&11&-0.3&-4&-3.2&-0.8&-5.9&-6&-7.6&1.6&-2.6&4.4&1.6&-4&1.9&-4.1\\
		\midrule
		qubit (sherbrooke) &{\bf 65}&69&{\bf 75}&{\bf 78}&{\bf 82}&86&89&{\bf 95}&99&{\bf 103}&107&113&116&{\bf 119}&123&126\\
		drive freq. [GHz]&4.758&4.837&4.769&4.861&4.815&4.89&4.948&4.802&4.832&4.695&4.986&4.964&4.931&4.793&4.821&4.831\\
		gate error $[10^{-4}]$&1.5&3.1&3&2.1&2.9&2.1&4.9&4.4&2.6&1.6&2.6&2.1&1.1&1.2&1.4&8.5\\
		$W^{i}[10^{-4}]$&{\bf 10}&-3.9&{\bf 7}&{\bf 11}&{\bf 9.3}&6.8&2.6&{\bf 15}&-3.1&{\bf 11}&1.4&-4.6&7.7&{\bf 15}&-3.2&7.6\\
		$\sigma^i[10^{-4}]$&{\bf 1.3}&1.4 &{\bf 1.2}&{\bf 1.3}&{\bf 1.4}&1.4&1.3&{\bf 1.4}&1.4&{\bf 1.4}&1&1.4&1.4&{\bf 1.3}&1.4&1.1
		 \\
		$W^{ii}[10^{-4}]$&{\bf 10}&-3.9&{\bf 7}&{\bf 11}&{\bf 9.3}&6.8&3&{\bf 15}&-3.1&{\bf 11}&1.3&-4.6&7.7&{\bf 15}&-3.3&7.7\\
		$\sigma^{ii}[10^{-4}]$&{\bf 1.3}& 1.4&{\bf 1.2}&{\bf 1.3}&{\bf 1.4}&1.4&1.3&{\bf 1.4}& 1.4&{\bf 1.4}& 1.1&1.4&1.4&{\bf 1.3}&1.4&1.1\\
		$p_{10}-p_{14}[10^{-4}]$&8.9&-3.7&4.4&0.6&2.8&-0.2&7.8&0.8&-9.8&-7.1&-8&7.5&-1.3&1.7&0.4&7.9\\
		$p_{11}-p_{15}[10^{-4}]$&-0.4&-1.3&0.0&-2.1&-2&1.6&-2.2&0.4&-1.6&-2.4&-1.1&6.1&6.6&-4.7&4.9&-5.9\\
		$p_{20}-p_{24}[10^{-4}]$&-2.9&-3.5&0.2&-4.7&-1.3&-1.1&-1.2&-0.8&1&-2.4&-4.4&3.6&2.2&-1.1&-6.9&7.7\\
		\bottomrule
	\end{tabular}
	\caption{Experimental data. Top: \emph{ibm\_brisbane} (April 2024),
		 bottom: \emph{ibm\_sherbrooke} (May 2024) on specified qubits.
		The data contain drive frequency (inter-level), error of the $\hat{S}$ gate,
		 the witness $W^{i/ii}$ and the standard deviation $\sigma^{i/ii}$ 
		for the two ways of averaging, explained in the text. Faulty qubits (beyond 5 standard deviations) are
		 bolded. The differences $p_{10}-p_{14}$, $p_{11}-p_{15}$, and $p_{20}-p_{24}$ should be zero for ideal gates. Here they indicate the range of empirical error of the $S$ gate.}
	\label{data}
\end{table*}

\section{Discussion}

The presented null test of qubit dimension turned out to fail in about a half of the tested qubits.
Due to the robustness of the test against common sources of error such as decoherence and gate instability, the deviations indicate either 
a serious device malfunction or resorting to more fundamental explanations. As the simulated leakage to higher excited states is $<10^{-5}$ \cite{leak},
only intricate technical imperfections may be the cause, in particular (i) memory effects  (residual qubit population or gate pulses) (ii)
parasitic transition in the neighbor qubits (iii) malfunctioning classical part (software, cables, pulse controls). In any case, these reasons cannot be deduced
simply from calibration data supplied by IBM Quantum. Otherwise, fundamental explanations
involving extra states beyond simple models predicting extra dimensions, as many worlds/copies \cite{plaga,abadp}, must be considered. 
One can generalize the tests  also in various directions, higher dimensions, 
entangled states, more complicated operations, or different platforms. Summarizing, the unprecedented accuracy of the test, and the ability to
identify the faulty qubits relatively quickly, demonstrates the extreme importance of certifying the building blocks of a quantum computer.

\section*{Acknowledgments}

The results have been created using IBM Quantum. The views expressed are those of the authors and do not reflect the official policy or position of the IBM Quantum team. We thank Jakub Tworzyd{\l}o for advice, technical support, and discussions, and Witold Bednorz for consultations on error analysis.
TR gratefully acknowledges the funding support by the
program ,,Excellence initiative research university'' for the AGH University in
Krakow as well as the ARTIQ project: UMO-2021/01/2/ST6/00004 and
ARTIQ/0004/2021. We also thank Bart{\l}omiej Zglinicki and Bednorz family for the support.

\appendix
\section{Estimation of leakage}

According to Qiskit documentation and IBM Quantum data, the gate $\hat{S}$ is realized by a time-dependent Hamiltonian
based on anharmonic oscillator \cite{gam1,gam2},
including the higher excited state $|2\rangle$,
\begin{equation}
\hat{H}=\Omega(t)(|1\rangle\langle 0|+\lambda|2\rangle\langle 1|)+\mathrm{h.c.}+\Delta|2\rangle\langle 2|,
\end{equation}
where $\Delta$ is the anharmonicity gap, $\Omega$ is the control pulse with real and imaginary parts, and $\lambda$ is the transition amplitude. We shall assume
$\lambda=\sqrt{2}$ as in harmonic oscillator framework. 
A truncated Gaussian pulse realizing $S$ gate has the form
\begin{equation}
\Omega(t)=\Omega_G(t)-\lambda^2i\dot{\Omega}_G(t)/4\Delta,
\end{equation}
with
\begin{equation}
\Omega_G(t)=(\pi/2)\frac{\exp(-t^2/2\sigma^2)-\exp(-T^2/8\sigma^2)}{\sqrt{2\pi}\sigma\mathrm{erf}(T/2\sqrt{2}\sigma)-T\exp(-T^2/8\sigma^2)},
\end{equation}
for $t\in [-T/2,T/2]$, with $T$ -- gate time, $\sigma$ -- width.
IBM Quantum gates operate with parameters, $T=n_T \delta$, $\sigma=n_\sigma \delta$, at $n_T=256$, $n_\sigma=64$, with the sampling time $\delta=0.222$ns
and anharmonicity $\Delta=2\pi \nu$ for $\nu=-310$MHz.
The imaginary part is responsible for Derivative Removal by Adiabatic Gate (DRAG) and we shall treat it as a correction
It can be understood if one considers the slow limit of the $1\leftrightarrow 2$ transition.

For a pefrect transition (in the limit $\Delta\to \infty$), the total evolution reads,
\begin{align}
&\hat{U}(t)=\exp\int_{-\infty}^t\Omega_G(t')\sigma_{x,0,1}dt'/i=\nonumber\\
&\begin{pmatrix}
\cos\phi(t)&-i\sin\phi(t)\\
-i\sin\phi(t)&\cos\phi(t)
\end{pmatrix},
\end{align}
in $|0\rangle$, $|1\rangle$ basis 
with 
\begin{equation}
\phi(t)=\int_{\infty}^{t}\Omega_G(t')dt',
\end{equation}
and $\sigma_{x01}=|1\rangle\langle 0|+|0\rangle\langle 1|$. Now, for the real part of $\Omega$,
probability of the transition to the state $|2\rangle$ of the initial state $|\psi\rangle=\psi_0|0\rangle+\psi_1|1\rangle$ is
\begin{equation}
p(\psi\to 2)\simeq 2\left|z\psi_1-iz^\ast \psi_0\right|^2,
\end{equation}
with
\begin{align}
&z=\int_{-T/2}^{T/2}  e^{i\Delta t}\Omega_G(t)\cos\phi(t)dt\nonumber\\
&=e^{i\Delta T/2}-i\Delta\int_{-T/2}^{T/2} e^{i\Delta t}\sin\phi(t)dt.
\end{align}
The maximum is for $\psi_0/\psi_1=i(z/z^\ast)^{1/2}$, i.e. $4|z|^2$, equal $3.6\cdot 10^{-7}$ for IBM Quantum data.
The imaginary part counteracts only the effect of phase shift within the $0-1$ space, caused by a \emph{temporary} leakage
during the pulse.
To understand it, let us expand the correction to $U\equiv U(T)$,
\begin{align}
&\Delta \hat{U}=i\hat{U}\int dt \hat{U}^\dag(t)\lambda^2\dot\Omega_G(t)\sigma_{y01}\hat{U}(t)/4\Delta\nonumber\\
&-\hat{U}\int da \hat{U}^\dag(a)\Omega_G(a)h(a,b)\Omega_G(b)\lambda^2|1\rangle\langle 1| \hat{U}(b),
\end{align}
with $\sigma_{y01}=i|1\rangle\langle 0|-i|0\rangle\langle 1|$.
Let us consider the Fourier transform of 
\begin{equation}
h(a,b)=e^{i\Delta(b-a)}\theta(a-b),
\end{equation}
which is
\begin{align}
&\int e^{i\alpha a+i\beta b}h(a,b)dadb=2\pi\delta(\alpha+\beta)\int dt e^{i\alpha t-i\Delta t}\theta(t)dt\nonumber\\
&=
\frac{2\pi\delta(\alpha+\beta)}{\epsilon+i\Delta-i\alpha}.
\end{align}
For small $\alpha$ we can expand
\begin{equation}
\simeq 2\pi\delta(\alpha+\beta)/i\Delta,
\end{equation}
which is a Fourier transform of
\begin{equation}
\tilde{h}(a,b)=\delta(a-b)/i\Delta.
\end{equation}
Replacing $h$ with $\tilde{h}$  and, integrating by parts the first term, we have
\begin{align}
&\Delta \hat{U}\simeq -\int dt \dag{U}(t)\Omega^2_G(t)[\sigma_{x01},\sigma_{y01}]\hat{U}(t)\lambda^2/4\Delta\nonumber\\
&=-\hat{U}\int  \hat{U}^\dag(t)\Omega^2_G(t)\lambda^2|1\rangle\langle 1| \hat{U}(t)/i\Delta\nonumber\\
&
\simeq -U\int idt \dag{U}(t)\Omega^2_G(t)(\hat{I}_{01})]\hat{U}(t)\lambda^2/2\Delta,
\end{align}
where $\hat{I}_{01}=|0\rangle\langle 0|+|1\rangle\langle 1|$, identity in $0-1$ space.
Finally
$
\hat{U}\to \hat{U}(1+i\theta),
$
where
$
\theta=\int \Omega^2_G(t)\lambda^2dt/2\Delta
$
is an unobservable global phase shift.

\end{document}